\documentclass[amsmath,amssymb,aps,preprint,floatfix,superscriptaddress]{revtex4-1}  
\usepackage{graphicx}
\usepackage{stmaryrd}
\usepackage{dcolumn}
\usepackage{bm}
\usepackage[usenames,dvipsnames]{xcolor}
\usepackage{multirow}
\usepackage{times}
\usepackage{chngcntr}
\usepackage{hyperref}
\usepackage{lineno} 

\makeatletter
\newcommand{\mat}[1]{{\mathpalette\mat@{#1}}}
\newcommand{\mat@}[2]{%
  \begingroup
  \sbox\z@{$\m@th#1\underline{#2}$}%
  \dimen@=\dp\z@ \advance\dimen@ -2\mat@dimen{#1}%
  \dp\z@=\dimen@
  \sbox\z@{$\m@th\underline{\box\z@}$}%
  \box\z@
  \endgroup
}
\newcommand\mat@dimen[1]{%
  \fontdimen8
  \ifx#1\displaystyle\textfont\else
  \ifx#1\textstyle\textfont\else
  \ifx#1\scriptstyle\scriptfont\else
  \scriptscriptfont\fi\fi\fi 3
}
\makeatother

\newcommand{\vc}[1]{\ensuremath{\boldsymbol{#1}}}

\newcommand{\code}[1]{{\normalfont\ttfamily #1}}
\newcommand{\unit}[1]{\mathrm{\,#1}}
\newcommand{\EF}{\ensuremath{E_\mathrm{F}}}

\usepackage[normalem]{ulem} 

\usepackage{placeins}

\begin{document}


\title{Proximity induced superconductivity in a topological insulator}

\author{Philipp R\"u{\ss}mann} 
\email[Corresponding author: ]{p.ruessmann@fz-juelich.de}
\affiliation{Institute of Theoretical Physics and Astrophysics, University of D-97074 Würzburg, Germany}
\affiliation{Peter Gr\"unberg Institut and Institute for Advanced Simulation, 
	Forschungszentrum J\"ulich and JARA, D-52425 J\"ulich, Germany}
\author{Stefan Bl\"ugel} 
\affiliation{Peter Gr\"unberg Institut and Institute for Advanced Simulation, 
	Forschungszentrum J\"ulich and JARA, D-52425 J\"ulich, Germany}
\date{\today}


\begin{abstract}
    Interfacing a topological insulator (TI) with an $s$-wave superconductor (SC) is a promising material platform that offers the possibility to realize a topological superconductor through which Majorana-based topologically protected qubits can be engineered. 
    In our computational study of the prototypical SC/TI interface between Nb and Bi$_2$Te$_3$, we identify the benefits and possible bottlenecks of this potential Majorana material platform.
    Bringing Nb in contact with the TI film induces charge doping from the SC to the TI, which shifts the Fermi level into the TI conduction band.
    For thick TI films, this results in band bending leading to the population of trivial TI quantum-well states at the interface.
    In the superconducting state, we uncover that the topological surface state experiences a sizable superconducting gap-opening at the SC/TI interface, which is furthermore robust against fluctuations of the Fermi energy.
    We also show that the trivial interface state is only marginally proximitized, potentially obstructing the realization of Majorana-based qubits in this material platform.
\end{abstract}

\maketitle


\section{Introduction}
\label{sec:intro}

The promise to realize topologically protected qubits based on the non-Abelian exchange statistics of Majorana zero modes that can be nucleated in topological superconductors sparked a lot of research interest in the field in the past years~\cite{Fu2008, Cook2011, Alicea2011, Sato2017, Flensberg2021}.
Despite the ongoing effort in increasing, for example, the decoherence times of qubits, the entire field of quantum computing still faces material optimization challenges of the different pursued qubit architectures and Majorana-based platforms are no exception~\cite{DeLeon2021}.

TIs are a very versatile material and recent advances in fabrication techniques open up new ways to use their unique properties for novel functionalities in devices~\cite{Hasan2010, Breunig2022}. For example, recently a TI-based Josephson junction was fabricated that demonstrated the usability of TIs in the field of superconducting transmon qubits~\cite{Schmitt2022}. The fundamental property of the superconducting proximity effect in SC/TI heterostructures has been a major research focus in the past years, which was pursued both experimentally~\cite{Xu2015, Xu2014, Stolyarov2021, Flototto2018, XuAlidoust2014, Bai2020, Schmitt2022} and theoretically~\cite{Fu2008, Chiu2016, Park2020, Legg2022}. For instance, using scanning tunneling microscopy~\cite{Xu2015, Xu2014, Stolyarov2021}, angle-resolved photoemission~\cite{Flototto2018, XuAlidoust2014}, or transport experiments~\cite{Bai2020}, it was demonstrated that TIs of the (Bi,Sb)$_2$(Te,Se)$_3$ family can indeed be proximitized by an interface to a superconductor. Elemental $s$-wave superconductors like Nb~\cite{Schmitt2022, Flototto2018} or Pb~\cite{Stolyarov2021} but also two-dimensional van der Waals superconductors like NbSe$_2$~\cite{Xu2015, Xu2014, XuAlidoust2014} or PdTe$_2$~\cite{Park2020, Bai2020} have been used successfully to open a superconducting gap in the electronic structure of a TI.
It was demonstrated that the proximity effect decays exponentially in the topological surface state (TSS) with its distance to the SC/TI interface (i.e.\ with the TI film thickness)~\cite{Flototto2018, Xu2014, Park2020}. Further analysis even revealed possible signatures of Majorana zero modes in form of zero-bias voltage peaks inside supercurrent vortices of the SC/TI heterostructures~\cite{Xu2015, Xu2014}.

Comparing different SC/TI interfaces, it is clear that the details of the interface chemistry strongly influence the proximity effect. This is reflected (i) in the different extent to which the quantum-well states and the TSS of thin TI films are gapped out upon contact with a SC and (ii) in how quickly the proximity-induced gap in the different states decay from the SC/TI interface~\cite{XuAlidoust2014, Flototto2018, Park2020}. These differences in the proximity effect were attributed to the degree of hybridization of the respective TI wave functions with the superconductor and the resulting possibility for Cooper pair tunneling into the TI~\cite{XuAlidoust2014, Park2020}.
Only in the recent past, it was realized that at the SC/TI contact charge transfer and band bending effects play an important role for the proximity effect and the opening of a topological gap~\cite{Stolyarov2021, Legg2022}. It was understood that the effect of charge transfer can even be beneficial in proximitized TI nanowires due to the induced inversion symmetry breaking, as it can enhance the sub-band splitting of TI states required for the realization of topological superconductivity in this geometry~\cite{Legg2022}.

Despite the numerous research on the proximity effect in SC/TI heterostructures, the interplay of the microscopic details of the interface, the local electronic structure, and the factors controlling the proximity effect in this material platform remains elusive, and to date no topological qubit exists. Thus, to fill this gap, a deeper understanding of the physics of the SC/TI interface is obligatory, which we aim at in this study.

The focus of this computational study is the induced proximity effect in the electronic structure of the TI Bi$_2$Te$_3$ in contact to the $s$-wave superconductor niobium (Nb).
For the TI we consider thin films of Bi$_2$Te$_3$ of different thicknesses between two and ten quintuple layers (QLs), which corresponds to thicknesses of $\approx 2 - 10$\,nm. The SC contact is modelled using six layers of Nb(111) in order to build a minimal structural model for the SC/TI interface.
Our calculations are based on density functional theory (DFT) and the Bogoliubov-de Gennes (BdG) method to include the effect of superconductivity together with a realistic description of the electronic structure~\cite{RuessmannKKRBdG}. Further calculational details can be found in the ``Methods'' section.

The calculational setup is shown in Fig.~\ref{fig:setup}(a) where the ``free'' (i.e.\ the QL adjacent to the vacuum region) and the ``contact'' QL (i.e.\ the QL in contact to the superconductor) are highlighted. Throughout this work we discuss the electronic structure on both free and contact sides of the TI film. In general, in the TI/SC heterostructure a TSS will appear on the free side at the interface between TI and vacuum, and also as a contact TSS at the interface between the TI and the SC which can hybridize with the states from the SC. Chemically, the SC/TI interface is a metal-semiconductor contact where charge transfer and band bending effects can lead to the occupation of trivial interface states in the accumulation region~\cite{Bahramy2012, Michiardi2022}. Our work discusses these different states that appear at the SC/TI contact and uncovers the interplay of wave function localization, superconducting proximity effect which may allow to further optimize the SC/TI interface using elemental SCs in the future.


\section{Results}

\subsection{Normal state properties of the Bi$_2$Te$_3$/Nb interface}

\begin{figure*}
    \centering
    \includegraphics[width=\linewidth]{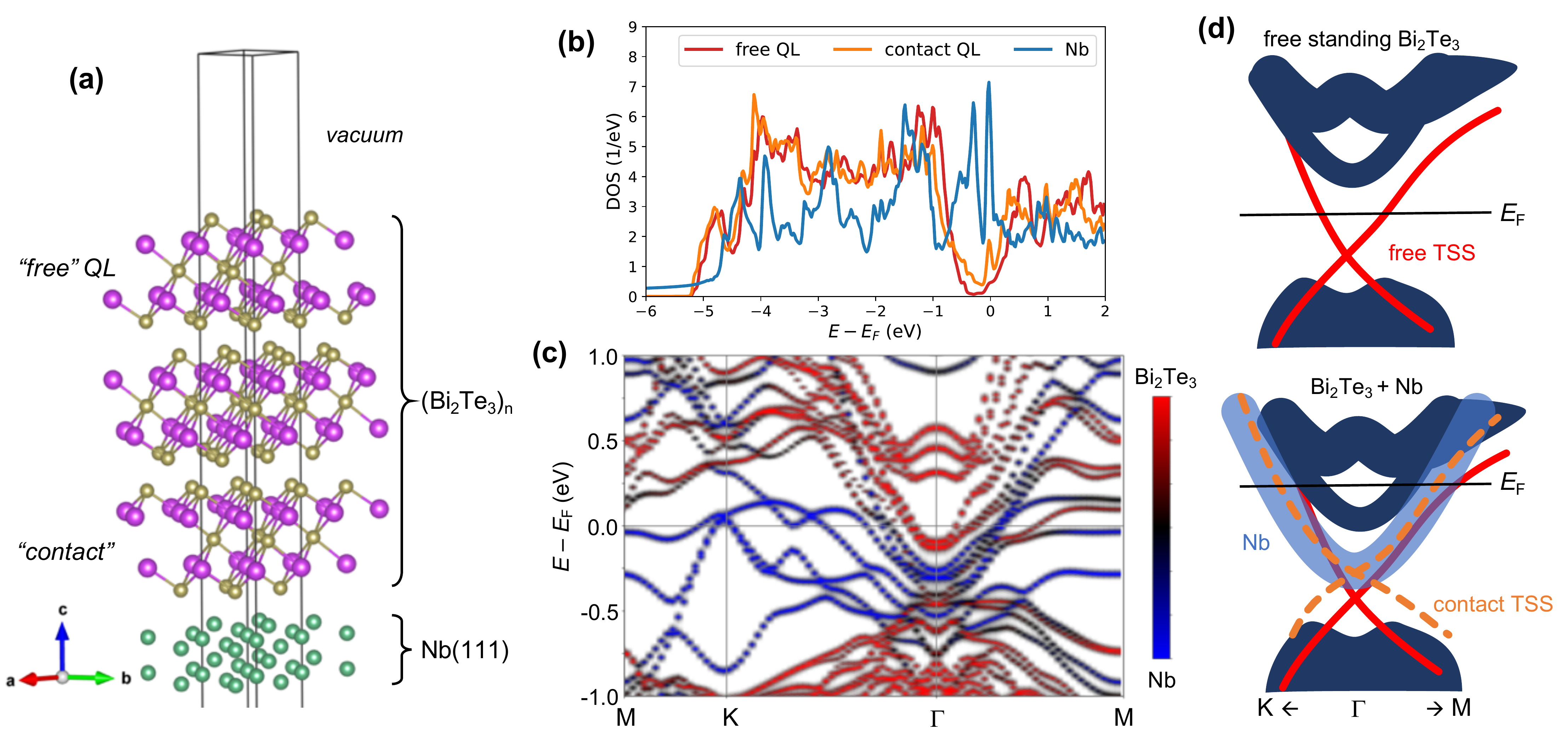}
    \caption{Calculational setup and normal-state electronic structure of a topological insulator -- superconductor (SC/TI) heterostructure. \textbf{(a)} Calculational setup of $n$ quintuple layers (QLs) Bi$_2$Te$_3$ ($n=2-10$) in contact with 6 layers Nb(111). Indicated are the ``contact'' and the ``free'' QLs of the Bi$_2$Te$_3$ film. \textbf{(b)} Density of states integrated in the regions of the free QL, the contact QL and the Nb region for the 3 QL SC/TI heterostructure shown in (a).
    \textbf{(c)} Electronic band structure of the SC/TI heterostructure where the localization of the wave functions is given as red (localized in Bi$_2$Te$_3$) or blue (localized in Nb) colors. The Dirac point of the topological surface state (TSS) on the free Bi$_2$Te$_3$ side is seen around $E\approx-0.45\unit{meV}$ at the $\Gamma$-point. \textbf{(d)} Schematic representation of the changes to the TSS upon bringing the TI in contact to Nb.}
    \label{fig:setup}
\end{figure*}

Figure~\ref{fig:setup}(b) shows an exemplary normal state density of states (DOS) without superconductivity for a three QL thick Bi$_2$Te$_3$ film in contact to Nb. The DOS is integrated over the regions of the free QL, the contact QL and the Nb region. The DOS in the free QL hardly changes compare to the DOS in freestanding Bi$_2$Te$_3$ (not shown). In contrast, the QL in contact with Nb clearly hybridizes with the states originating from the Nb film, introducing additional states that are particularly visible in the region of the TI bulk band-gap around $E-\EF=-0.3\unit{eV}$.
Contacting Nb with Bi$_2$Te$_3$ leads to a charge transfer from Nb to the TI of around $0.15$ electrons ($\approx0.87$ electrons per nm$^{2}$), resulting in an upward shift of the Fermi level. This Fermi-level shift is obvious from the band structure shown in Fig.~\ref{fig:setup}(c) where the color code indicates the localization of the states. The  Fermi level resides inside the conduction band of the TI as shown schematically in Fig.~\ref{fig:setup}(d). Furthermore, the hybridization of the TSS in the contact QL with states from the Nb film strongly affects the dispersion of the surface state, which is visible in an upward shift and an opening of the Dirac cone formed by the TSS.

\subsection{Superconducting proximity effect in ultra-thin SC/TI heterostructures}

\begin{figure*}
    \centering
    \includegraphics[width=0.45\linewidth, trim={0.0cm 0.2cm 0cm 0.2cm}, clip]{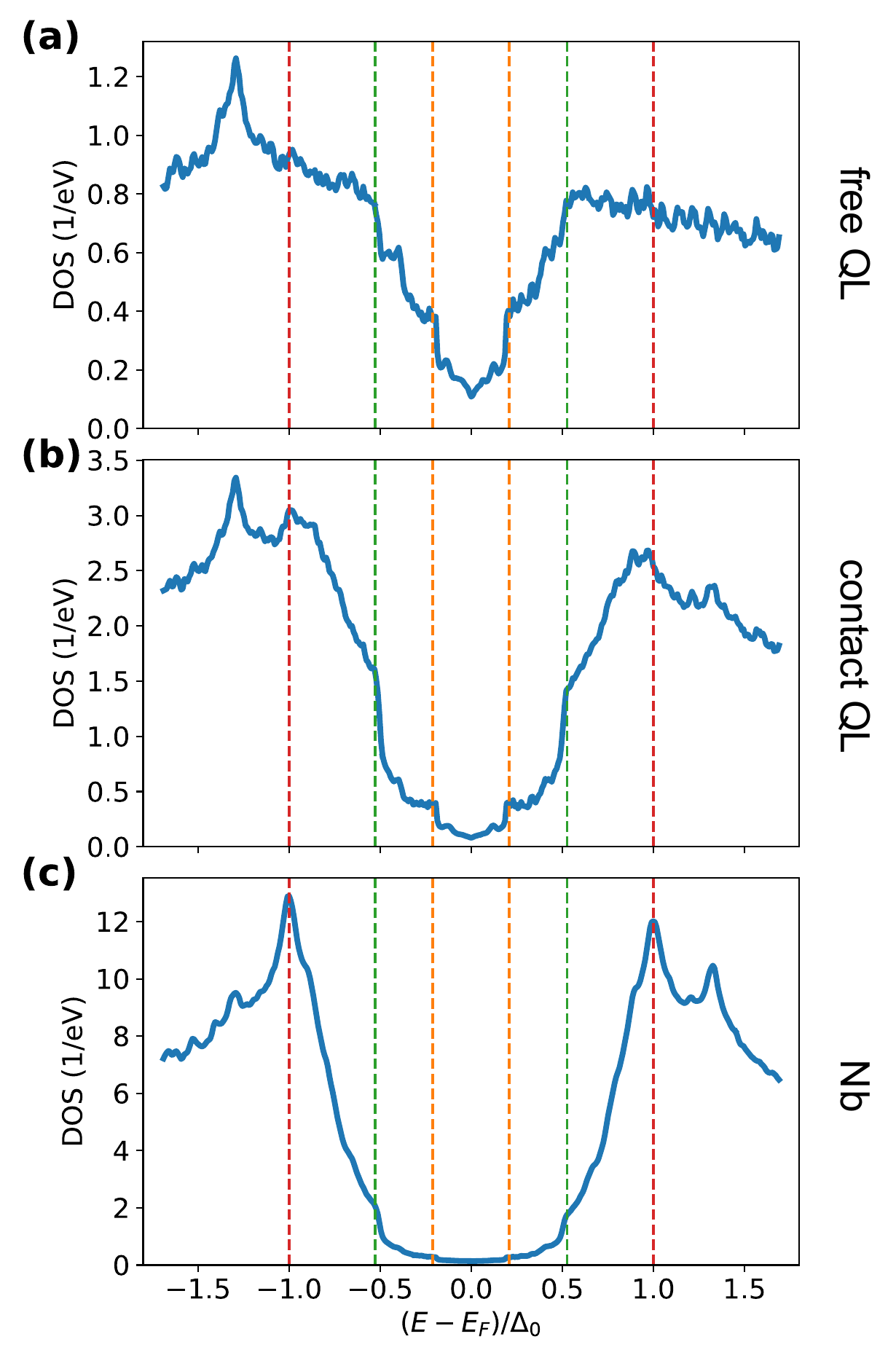}
    \caption{Superconducting density of states for 2 QLs Bi$_2$Te$_3$ on Nb(111). \textbf{(a-c)} integrated DOS in the free QL, the contact QL and the Nb region, respectively. Three distinct gap sizes seen in the DOS are indicated with dashed vertical lines which highlight the superconducting gap of Nb (red, $E=\pm \Delta_0$) the principal induced gap in the contact quintuple layer (green, $E=\pm0.53\Delta_0$) and the induced gap in the free QL (orange, $E=\pm 0.21\Delta_0$). Shown is only the particle component of the particle-hole space used in the BdG formalism.}
    \label{fig:BdGdos}
\end{figure*}

We start the investigation of the superconducting proximity effect in the topological insulator with the thinnest TI film we consider here. For two QLs Bi$_2$Te$_3$ the free and the contact side are not completely decoupled as seen, for example, in the sizable hybridization-induced gap at the Dirac points in the TSS of freestanding TIs in the ultra-thin limit \cite{ZhangTSSgap}.
Figure~\ref{fig:BdGdos} shows the density of states in the superconducting state of the two QLs of Bi$_2$Te$_3$ and the Nb regions in a narrow energy window around the Fermi level. The superconducting $s$-wave pairing in Nb leads to the well-known opening of a gap in the density of states of the Nb layers. The principal coherence peak of Nb is used to identify the size of the superconducting gap in the superconductor region (see Fig.~\ref{fig:BdGdos}c) which is called $\Delta_0$. The rich structure in the electronic bands at the complex SC/TI interface, together with the hybridization of the states and their partial localization in Nb or the TI region, weaken the superconductivity at the interface and lead to a soft superconducting gap where the DOS does not abruptly drop to zero for energies $|E-E_\mathrm{F}|<\Delta_0$.
A second gap at roughly half the size of intrinsic Nb gap (highlighted with green dashed lines in Fig.~\ref{fig:BdGdos}b) arises from the contact QL of the Bi$_2$Te$_3$ film. This proximity induced gap is sizable due to the strong overlap of the wave functions in the interface regions. Finally, in the DOS of the free QL a third gap is evident which reaches only a fifth of the value of the Nb gap (orange dashed lines). We attribute this to the surface states arising on the free side that only has an exponential tail of its wave functions extend into the superconductor region. Therefore, the proximity induced gap in the free side is significantly reduced compared to the contact side.

\begin{figure*}
    \centering
    \includegraphics[width=0.7\linewidth]{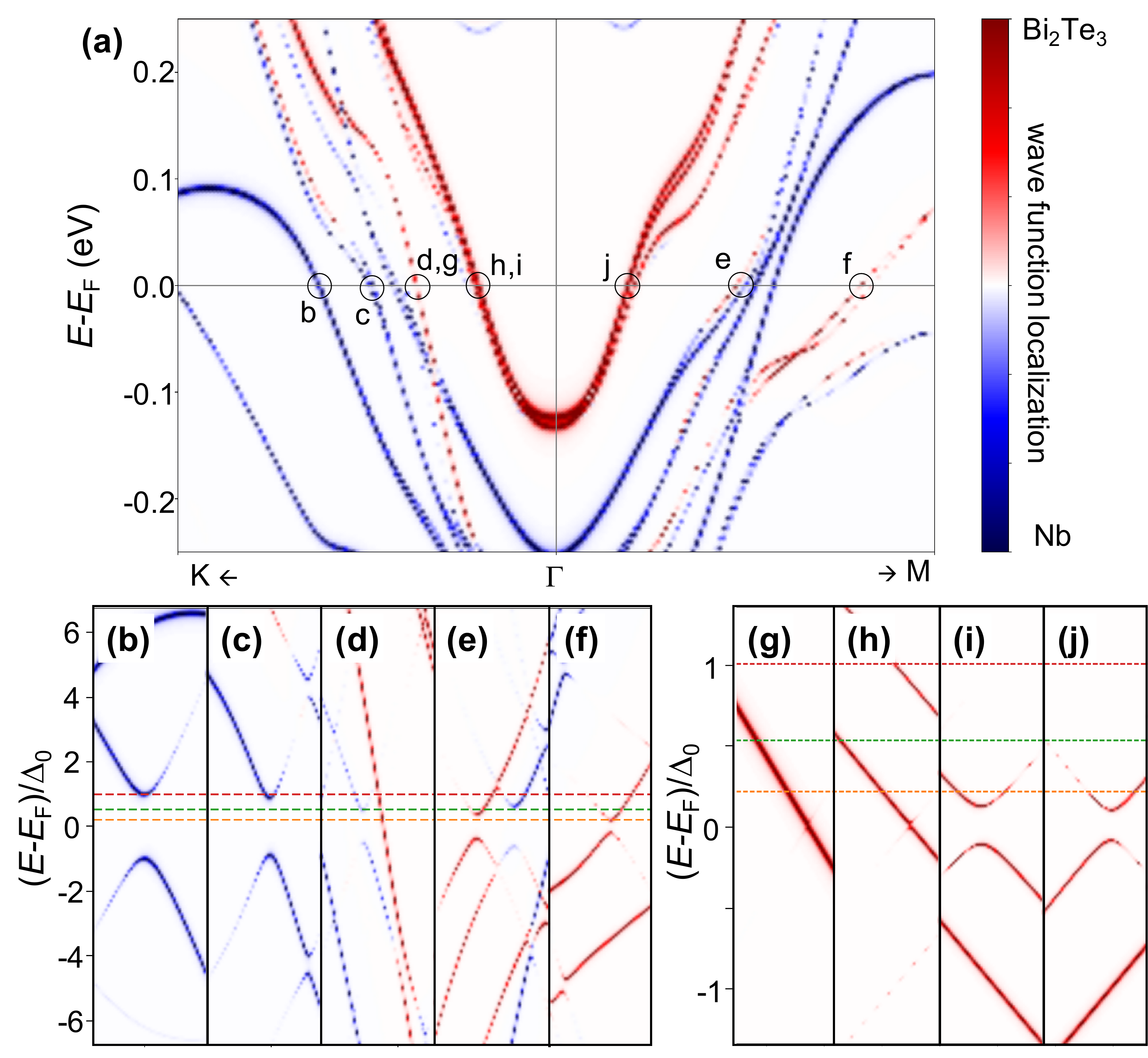}
    \caption{Superconducting gap in different bands of a 2QL Bi$_2$Te$_3$ / Nb(111) heterostructure. \textbf{(a)} Normal state band structure around the $\Gamma$ point. \textbf{(b-j)} Gap in the superconducting state of the bands around the intersections with the Fermi level marked in (a). The three dashed horizontal lines indicate the features in the density of states highlighted in Fig.~\ref{fig:BdGdos}. Note that the superconducting proximity gap in the topological surface state is shown in the red bands for $\overline{\Gamma K}$ on the free side in (d) and for $\overline{\Gamma M}$ on the free and the contact sides in (e) and (f), respectively.}
    \label{fig:2QLgapsBS}
\end{figure*}

Further insides into the proximity-induced gap on both sides of the TI film can be gained from an analysis of the superconducting band structure. Figure~\ref{fig:2QLgapsBS} shows the band structure of 2 QLs Bi$_2$Te$_3$ on Nb around the Fermi energy. The color code indicates the localization of the wave functions and panels (b-j) show magnified views into selected intersections of bands with the Fermi energy where the superconducting energy gap opens up. We recover the three identified gap sizes from the DOS plots in Fig.~\ref{fig:BdGdos} in states that are, respectively, mainly localized in Nb (blue bands in Fig.~\ref{fig:2QLgapsBS}b,c), states that are localized in the contact QL (panel e) or states mainly localized in the free side of the Bi$_2$Te$_3$ thin film (panels g,h).
We point out that the states at (d) and (e,f) arise from the TSS whereas the states at (h,i,j) are bulk-like quantum-well states of the thin TI films. 
In agreement with the DOS calculations discussed before, the induced gap in the TSS states in particular in $\overline{\Gamma K}$ is significantly smaller than the gap in the quantum-well states that couple much more strongly to the Nb states (panels i and j). Furthermore, it is noteworthy that the induced gap in the TSS in the orthogonal directions $\overline{\Gamma K}$ and $\overline{\Gamma M}$ differ considerably (cf.\ Fig.~\ref{fig:2QLgapsBS}f,g). 
The two states have a different Fermi velocity which is inversely proportional to the density of states at the Fermi energy $N(\EF)=\int_{\vc{k}\in\mathrm{FS}} (\hbar|\vc{v}_\mathrm{F}(\vc{k})|)^{-1}\,\mathrm{d}^2k$.
Consequently the superconducting gap size $\Delta \sim \exp \bigl( -\frac{1}{\lambda N(\EF)} \bigr)$, that follows from the BCS theory \cite{BCS} with the electron-phonon coupling constant $\lambda$, is expected to be larger for the crossing of the TSS with \EF\ in $\overline{\Gamma M}$ compared to the $\overline{\Gamma K}$ direction. Nevertheless, the localization of the wave function (the state in $\overline{\Gamma M}$ has a longer wave function tail into the Nb region) will also play a significant role in the formation of the differently sized induced gap in the TSS and cannot be neglected. 

\subsection{Band bending and superconducting proximity effect in thick TI / SC hetrostructures}

\begin{figure*}
    \centering
    \includegraphics[width=0.75\linewidth, trim={0cm 0.6cm 0cm 0cm}, clip]{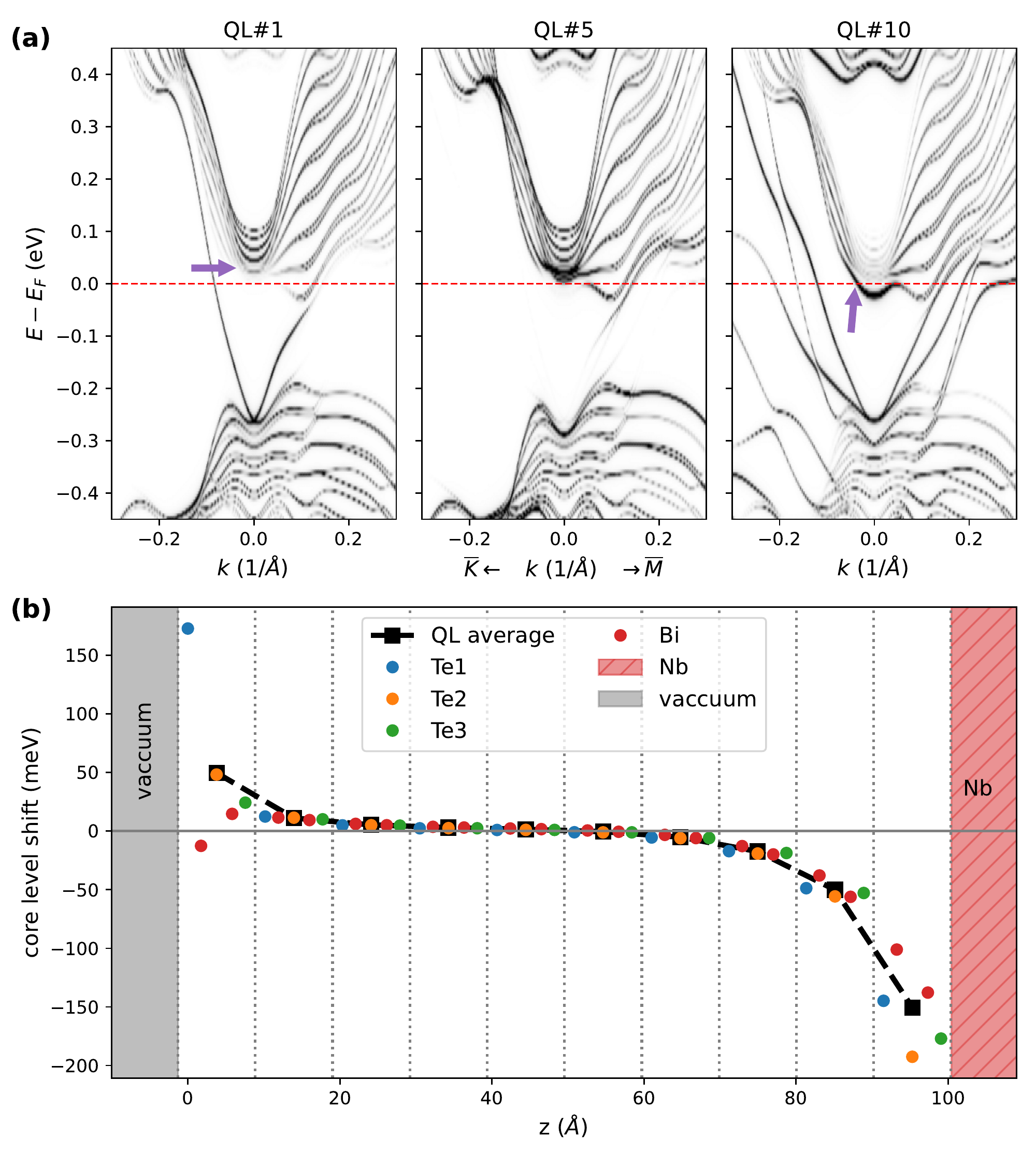}
    \caption{Band bending in $\sim 10\,\mathrm{nm}$ thick Bi$_2$Te$_3$ on Nb. \textbf{(a)} Band structure contributions of first, fifth and tenth (i.e.\ contact) quintuple layer (QL) which shows the band bending as a result of screening of the excess charge transferred from the Nb contact to the topological insulator (highlighted with purple arrows indicating the bottom of the conduction band around $\overline{\Gamma}$). \textbf{(b)} Core level shifts of the highest lying core state in the TI region (Te $4d$ and Bi $5d$). Shown are the contributions of the 3 distinct Te layer, the Bi layers as well as the average value over the QLs.}
    \label{fig:bandbending}
\end{figure*}

We now turn our attention to the limit of thick TI films in contact with the $s$-wave superconductor. Figure~\ref{fig:bandbending} shows the band structure of the 10 QL Bi$_2$Te$_3$ / Nb(111) heterostructure where the contribution of different QLs from the free side ({QL\#1}) to the contact side with Nb ({QL\#10}) are shown in (a). On the free side the Fermi level resides within the bulk band-gap below the onset of the conduction band marked with a purple arrow and only the TSS bridges the gap to the valence band. On the contact side on the other hand the charge transfer from the Nb contact shifts the Fermi level shifts upwards, which fills a state at the bottom of the conduction band (highlighted with the purple arrow). This effect is particularly strong in the topological insulator because of the low density of states in the bulk band-gap that can screen the excess charge. As shown in the calculated shift of the highest lying core levels of the Bi and Te atom in Fig.~\ref{fig:bandbending}(b), the screening of the charge doping by the Nb contact decays over $\sim 4$ QLs after which the Fermi level also approaches the bottom of the conduction band (as seen in the middle panel of (a)). 

\begin{figure*}
    \centering
    \includegraphics[width=0.45\linewidth, trim={0cm 0.25cm 0cm 0cm}, clip]{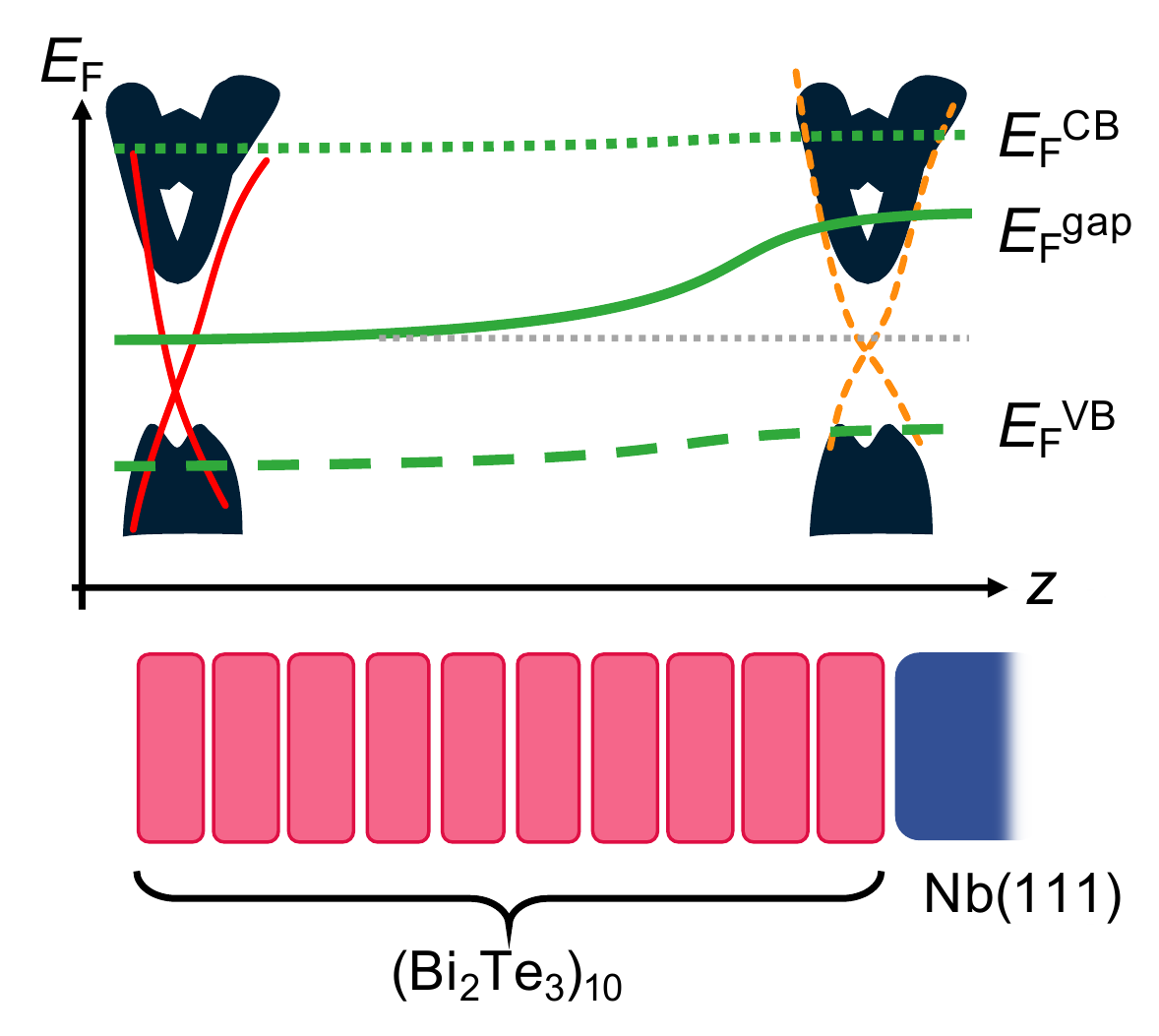}
    \caption{Schematic illustration of the charge-transfer-induced band bending in the TI region at different positions of the Fermi level located in the bulk band gap, the valence band (VB) or the conduction band (CB). Note that the electron transfer from Nb to the TI always leads to an upward shift of the Fermi level at the SC/TI contact which is screened more efficiently when the DOS in the TI region is higher (i.e.\ when $E_\mathrm{F}$ lies in VB or CB).}
    \label{fig:bandbendingSchematic}
\end{figure*}

\begin{figure*}
    \centering
    \includegraphics[width=0.7\linewidth, trim={0cm 0.0cm 0cm 0cm}, clip]{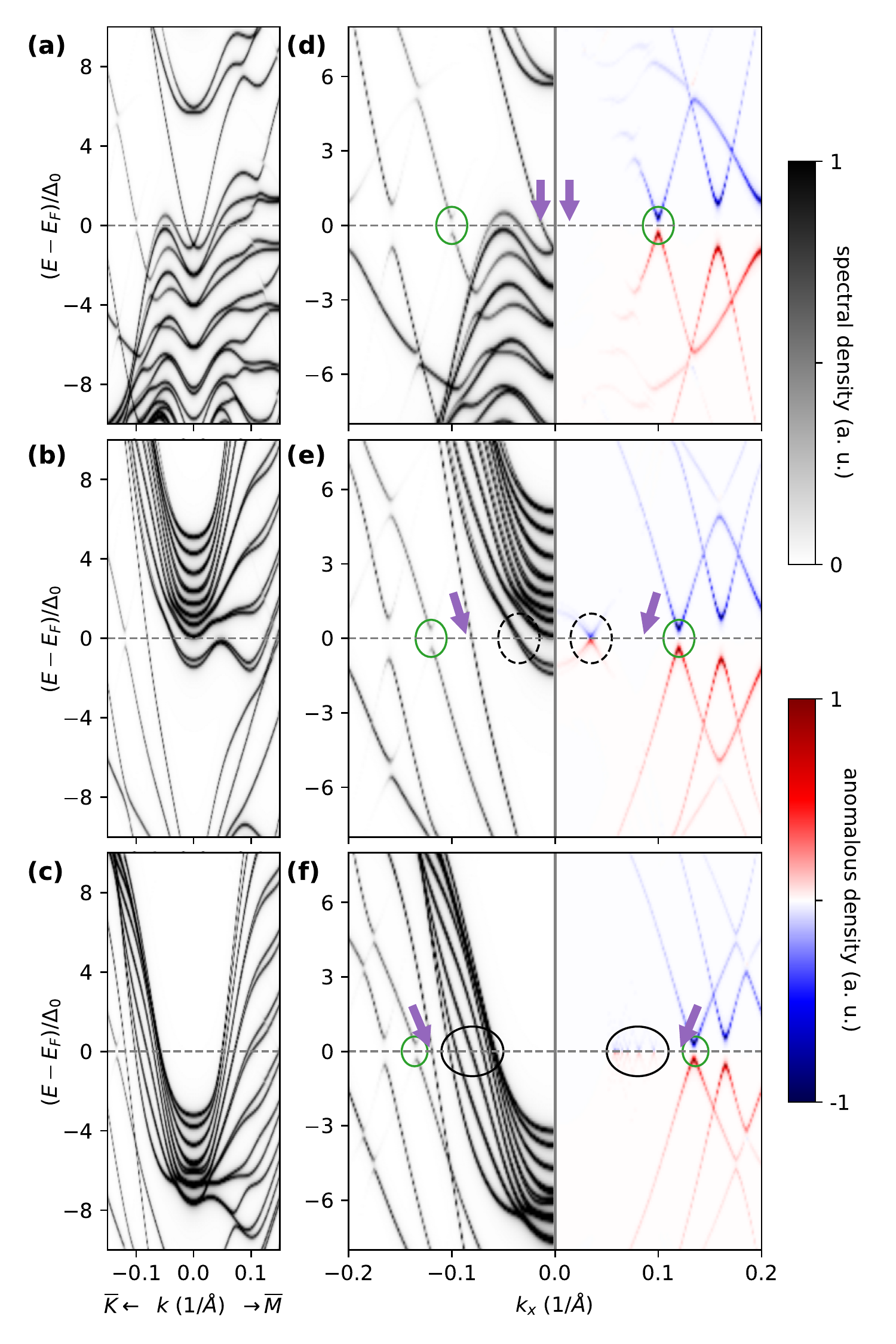}
    \caption{Superconducting band structure of 10 QL Bi$_2$Te$_3$ / Nb(111) for varying Fermi level in the bulk region of the TI changing from \textbf{(a)} Fermi level inside the bulk TI valence band, \textbf{(b)} inside the bulk band-gap and \textbf{(c)} inside the conduction band. \textbf{(d-f)} zoomed views of (a-c) along $\overline{\Gamma K}$ comparing the spectral density ($k_x<0$) and the anomalous density ($k_x>0$). The purple arrows (green ellipses) in (d-f) highlight the topological surface state on the free (contact) side of the SC/TI heterostructure.}
    \label{fig:bandstrucEFshift}
\end{figure*}

Next, we discuss the influence of a variation of the Fermi level on the band bending and the resulting superconducting gap in the electronic structure shown schematically in Fig.~\ref{fig:bandbendingSchematic}. This models the effect of intrinsic charge doping and allows to investigate the effect of charge puddles on the electronic structure and the proximity effect in SC/TI heterostructures. Figure~\ref{fig:bandstrucEFshift} shows the electronic structure in the normal state (a-c) compared to the superconducting state (d-f) for three locations of the Fermi level: (i) $E_\mathrm{F}$ inside the TI valence band, (ii) $E_\mathrm{F}$ inside the TI bulk band-gap as in Fig.~\ref{fig:bandbending}, and (iii) $E_\mathrm{F}$ inside the TI conduction band. Changing the location of the Fermi level in the TI region shifts the bands of the TI relative to the Nb bands which influences the hybridization of the TI states with the Nb states. Additionally, having the TI Fermi level located inside the valence or conduction band leads to higher density of states in the TI region when it comes in contact to the Nb so that the charge doping of from the Nb contact is screened efficiently and less band bending occurs.

The superconducting band structures are displayed in Fig.~\ref{fig:bandstrucEFshift}(d-f) for the three Fermi levels in the TI are shown in terms of the integrated spectral density (shown at $k_x<0$) that shows the particle-like components of the Bogoliubov-quasiparticles in $\overline{\Gamma K}$-direction. We further visualize the contributions to the anomalous density along the same path in the Brillouin zone ($k_x>0$) that indicates the amount to which different bands are proximitized. Here different bands pick up superconductivity due to the varying degree of tunneling of Cooper pairs from the superconductor into the respective TI bands because of the localization of the wavefunctions and their orbital character. Comparing the calculations for different Fermi levels, we find only minor changes in the superconducting gap of the bands that derive from the Nb layers ($|k_x|>0.15\,\mathrm{\AA}$) and in the TSS at the contact side (green ellipses). We conclude that the contact TSS hybridizes strongly with the superconductor and thus is proximitized efficiently with an induced gap size of roughly half of the gap of the superconductor. On the other hand the TSS on the free side is not proximitized as it is decoupled from the Nb contact through the thick TI film due to its exponential decay  from the free surface (see also Fig.~\ref{fig:wfloc} in the supplemental material).

Interestingly, for the Fermi level inside the TI bulk band-gap shown in Fig.~\ref{fig:bandstrucEFshift}(e), the filled state at the bottom of the conduction band is localized in the interface region with a maximum in the second QL from the Nb contact with only a small fraction of the wave function extending into the Nb region. This effectively decouples the interface state (highlighted with the dashed ellipses) from the superconductor so that the resulting proximity effect is hardly visible. Finally, the conduction band states of the TI, that cross the Fermi level in Fig.~\ref{fig:bandstrucEFshift}(f) in the region highlighted by the black ellipse, are located within the bulk of the TI film and thus are not proximitized at all.

\subsection{Decay of the superconducting proximity effect in the TSS}

\begin{figure*}
    \centering
    \includegraphics[width=0.75\linewidth]{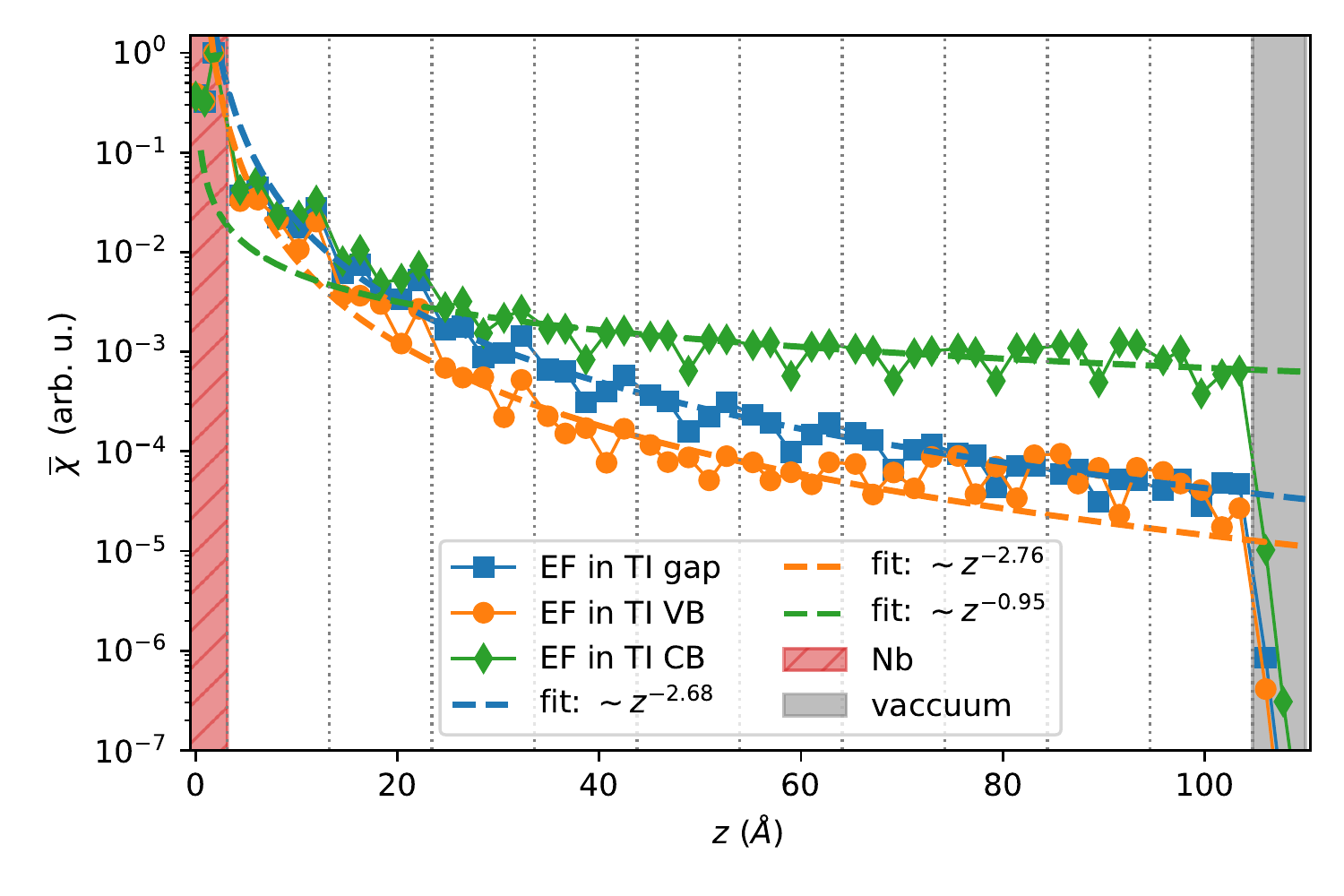}
    \caption{Decay of the superconducting order parameter (i.e.\ the average anomalous density) for 10 QL thick Bi$_2$Te$_3$ in contact to Nb. The shaded regions indicate the Nb and the vacuum regions and the vertical dotted lines highlights the beginning of the quintuple layers. The dashed lines show fits with functions $f(z) = \alpha z^{-\beta}$ where the fit exponent $\beta$ is given in the figure legend.}
    \label{fig:decayChi}
\end{figure*}

Finally, we discuss the decay of the superconducting proximity effect in the TI film. We focus on the results from the 10 QL thick TI film as the thickest film we considered in this work. Figure~\ref{fig:decayChi} shows the decay of the order parameter of superconductivity (i.e.\ the anomalous density $\chi$) from the Nb interface into the TI film. Shown are the result for the three positions of the Fermi level in the bulk TI region discussed in the previous section. Starting from the Nb layers on the left-hand side the order parameter quickly drops and characteristic oscillations in the atoms within the quintuple layers are seen. In the vacuum region the order parameter then drops exponentially as only the exponential tail of the TI wave functions penetrate into the vacuum.

In order to quantify the decay of the order parameter we fit a polynomial decay $f(z) = \alpha z^{-\beta}$ which we find to match our data much better than an exponential decay. In the fit we exclude the first two QLs as well as the vacuum region in order to analyze the long-distance limit without exaggerating effects from the region where band bending plays a major role. When the bulk TI Fermi level resides in the bulk band-gap or at the top of the valence band we find a fast decay ($\sim z^{-\beta}$) of the proximity effect in the superconducting order parameter with exponents $\beta=2.68$ and $\beta=2.76$, respectively. On the other hand the Fermi level lies within the conduction band of the TI, the decay of the order parameter approaches the slow decay of $\sim 1/z$ known from metal/SC interfaces~\cite{DeutscherdeGennes1969, Falk1963}. We attribute the faster decay of the order parameter for the Fermi level residing in the bulk band-gap or at the edge of the valence band with the significantly reduced number density of states in these circumstances which make the SC/TI interface behave rather like a semiconductor/SC interface than a metal/SC interface.

\section{Discussion}

Comparing our results to existing data in the literature, we find that the slow polynomial decay of the proximity effect in the order parameter is well in line with previous model-based studies of the proximity effect in SC/TI heterostructures~\cite{Chiu2016}.
Our results show that the localization of the wave functions in the SC/TI heterostructure determines the possibility for Cooper pair tunneling into the respective states. The TSS on the free side away from the SC contact is localized at the interface to the vacuum and only has an exponentially decaying tail of its wave function into the TI. Consequently the proximity gap in the free TSS shows an exponential decay with the TI thickness which is in line with earlier observations of the decaying proximity gap as seen for example in experiments~\cite{Xu2015, Xu2014, Stolyarov2021, Flototto2018, XuAlidoust2014}.

On the other hand, the absence of the proximity effect in some quantum-well states, which we similarly attribute to a negligible hybridization with Nb, is not always found experimentally~\cite{Flototto2018}. However, this apparent discrepancy could be an artifact of the structural model we chose in our computational approach. In general, the SC/TI interface between TIs and elemental superconductors is amorphous and poly-crystalline SC phases would be a more realistic description of the interface which is however computationally not feasible. A proper treatment would thus encompass extremely large unit cells with thick amorphous superconducting regions that show some intermixing at the SC/TI interface.
The defect-induced broadening of the states and the disordered nature of the interface can result in a quasi-continuum of states of SC states throughout the Brillouin zone with a slower decay into the surface TI region. Such a situation could therefore offer more possibilities for hybridization with the TI electronic structure which would result in more efficient Cooper pair tunneling into the TI and thus a larger induced proximity gap in the TI bands compared to our present results.

In contrast to amorphous interfaces with elemental SCs, the interfaces between vdW materials can in principle be much cleaner~\cite{Novoselov2016}, which will have consequences for the charge doping at the SC/TI interface. One can expect that the presence of the vdW gap between the different 2D materials (TI and SC) will lessen the charge doping which was also not reported in earlier studies using PdTe$_2$ and NbSe$_2$ to proximitize a TI \cite{Xu2014, XuAlidoust2014, Xu2015, Park2020}. Depending on the fabrication process a higher interface quality interface is however not a given as seen in the self-formed PdTe$_2$/(Bi,Sb)$_2$Te$_3$ heterostructure where considerable intermixing is present at the interface~\cite{Bai2020}.

In future studies, we plan to investigate the effect of intermixing and diffusion of superconductor atoms into the surface QLs of the TI and scrutinize its effect on the proximity effect. Such a dusting with a moderate concentration of superconducting atoms is believed to stabilize the proximity effect as shown in the very long observation of the superconducting gap amorphous atomic Pb layer on Bi$_2$Te$_3$ around Pb islands~\cite{Stolyarov2021}. We furthermore plan to probe the influence of buffer layers that can modify (e.g.\ partially screen) the charge doping from SC contact to the TI.

In conclusion, our results contain mixed messages for the ability to use elemental SC/TI interfaces as a platform for Majorana-based topological qubits. 
Encouraging is the sizable SC gap-opening we find in the contact TSS (in the order of half of the gap size of Nb), which is even robust against variations of the Fermi level in the TI region.
Therefore, the proximity effect in the TSS in contact with the SC is also expected to be robust to fluctuations in the Fermi level, as they occur in the presence of charge puddles, which are a common problem in TI materials~\cite{Beidenkopf2011, Skinner2012, Bomerich2017}.
Moreover, the charge transfer and the resulting band bending in thick TI films contacted with Nb are comparable with recent model calculations that showed the beneficial effect of metallization for the technologically important geometry of a proximitized TI nanowire~\cite{Legg2022}.
Detrimental for Nb/Bi$_2$Te$_3$ as a platform for topological superconductivity could be the trivial quantum-well states in the TI, which can become occupied due to band bending or intrinsic doping.
Depending on their localization at the SC/TI interface, the Cooper pair tunneling into these states is strongly suppressed, which leads to very small SC gap-openings in these trivial states.
Our findings may help in understanding and optimizing the interface between TIs and superconductors as a suitable material platform in the field of topologically protected quantum computing.


\section{Methods}

The density functional theory (DFT) results of this work are produced using the full-potential relativistic Korringa-Kohn-Rostoker Green's function method (KKR) \cite{Ebert2011} as implemented in the \code{JuKKR} code package~\cite{jukkr}. The local density approximation (LDA) is used to parameterize the exchange correlation functional \cite{Vosko1980}. We use $80\times80$ $k$-points in the Brillouin zone integration during self-consistency, spherical screening clusters of $1$~nm radius around each atom (containing $120-180$ sites), an $\ell_{max} = 2$ cutoff in the angular momentum expansion of the Green's function with an exact partitioning of the unit cell into atomic cells \cite{Stefanou1990,Stefanou1991}, and correct the truncation error arising from the finite $\ell_{max}$ cutoff using Lloyd's formula~\cite{Zeller2004}. A Lloyd's formula-like renormalization of the energy integration weights is used to shift the Fermi level of the topological insulator with respect to the Nb superconductor when the position of the Fermi level is changed in the calculation.
The TI/Nb(111) system in this study is calculated in a symmetric TI(111)/Nb(111)/TI(111) geometry in order to counteract spurious shifts in the electronic structure arising from charge-transfer-induced dipole moments. In order to facilitate the calculations we construct the TI/Nb interface starting from the experimental crystal structure of Bi$_2$Te$_3$ (111) of 4.4~\AA~\cite{ExpStrucBT}. Then 6 layers of Nb(111) scaled ($-6\%$ compared to Nb equilibrium lattice constant) to the same in-plane lattice constant as the TI film are placed in a distance of 2.6\,\AA~(i.e.\ the TI inter-layer distance). Because the focus of our investigation is the induced superconductivity in the electronic structure of the TI film we neglect the effect of lattice relaxations at the interface.

The superconducting properties are calculated based on the Bogoliubov-de Gennes (BdG) extension to the KKR method~\cite{Csire2015, RuessmannKKRBdG} which is abbreviated as ``KKR-BdG'' in this work. In the KKR-BdG method, the Bogoliubov-de Gennes Hamiltonian (given here in atomic units)
\begin{equation}
    H_{\mathrm{BdG}}(\vc{x}) =
      \left(
      \begin{array}{cc}
          -\nabla^2-E_{\mathrm{F}}+V(\vc{x}) & \mathcal{D}(\vc{x}) \\
          \mathcal{D}^*(\vc{x}) & \nabla^2+E_{\mathrm{F}}-V^*(\vc{x}) 
      \end{array}
      \right)    
\end{equation}
is solved self-consistently within the framework of density functional theory. Here, $\vc{x}$ denotes the position in space, $V(\vc{x})$ is the normal state electronic potential and $\mathcal{D}(\vc{x})$ is the superconducting pairing potential responsible for the formation of Cooper pairs. 
Following the parametrization of the exchange-correlation functional introduced by Suvasini \textit{et al.}~\cite{Suvasini1993}, the superconducting pairing potential in atom $i$ can be written as $\mathcal{D}_i(\vc{x})=\lambda_i \chi_i(\vc{x})$ with a set of atom-dependent semi-phenomenological coupling constants $\lambda_i$ and the anomalous density $\chi_i(\vc{x})$. In analogy to the BCS theory of superconductivity, the coupling constants $\lambda_i$ can be interpreted as the electron-phonon coupling, which results in the attractive interaction that form the Cooper pairs in conventional superconductors. We neglect the intrinsic electron-phonon coupling in the TI region and model the superconducting state with coupling constants
\begin{equation}
    \lambda_i = \left\{
        \begin{array}{rl}
             \lambda_0 & \mathrm{if}\ i \in \mathrm{sites\ of\ Nb\ atoms} \\
              0 & \mathrm{else}
        \end{array}
    \right.
\end{equation}
where $\lambda_0$ is a positive real-valued constant.
This is motivated by the smallness of the electron-phonon coupling in 3D TIs of the Bi$_2$Te$_3$-family that is found to be in the order of 0.1~\cite{Heid2017} whereas in Nb it is an order of magnitude larger~\cite{Bauer1998}.
The pairing potential in atom $i$ is calculated self-consistently using
\begin{equation}
    \mathcal{D}_i(\vc{x}) = \lambda_i \chi_i(\vc{x})
\end{equation}
with the anomalous density $\chi$ that is the order parameter of the superconductivity. This results in a finite intrinsic superconducting pairing potential in the Nb layers and vanishing intrinsic pairing potential in the TI layers.
More details of the KKR-BdG method can be found in Ref.~\onlinecite{RuessmannKKRBdG}.
In this work, we scale the coupling constant such that the superconducting gap in Nb is of the order of 1\,mRy to facilitate the resolution of the proximity-induced superconducting gaps in Bi$_2$Te$_3$. The induced gaps can then be scaled with respect to the experimental Nb gap size of 0.12~mRy (1.6~meV)~\cite{Odobesko2019}, where the scaling constant for the smaller gaps Bi$_2$Te$_3$ should remain the same.
We also varied the model for the coupling constant to stay constant throughout the heterostructure with the same value in the TI region and the SC region. This however does not affect the outcome of the results (i.e.\ the band structures and the decay of the anomalous density stay unaffected), which highlights the resilience of our conclusions with respect to this approximation.

The series of DFT calculations in this study are facilitated through the AiiDA-KKR plugin~\cite{aiida-kkr-code, aiida-kkr-paper} to the AiiDA infrastructure~\cite{aiida1}. This has the advantage that the data provenance is automatically stored in compliance to the FAIR principles of open data~\cite{Wilkinson2016}. The complete data set that includes the full provenance of the calculations is made publicly available in the materials cloud repository~\cite{Talirz2020, doi-dataset}.


\section*{Data Availability}

The data was generated using the JuKKR code (\url{https://jukkr.fz-juelich.de}) via the AiiDA-KKR plugin \cite{aiida-kkr-paper, aiida-kkr-code} to the AiiDA infrastructure~\cite{aiida1} and are publicly available on the materialscloud archive~\cite{Talirz2020, doi-dataset}.

\section*{Code Availability}

The source codes of the AiiDA-KKR plugin~\cite{aiida-kkr-paper, aiida-kkr-code} and the JuKKR code~\cite{jukkr} are published as open source software under the MIT license at \url{https://github.com/JuDFTteam/aiida-kkr} and \url{https://iffgit.fz-juelich.de/kkr/jukkr}, respectively.


\section*{Acknowledgements}

We thank Achim Rosch and Henry Legg for fruitful discussions about the effect of band bending in the SC/TI heterostructure. We acknowledge support by the Joint Lab Virtual Materials Design (JL-VMD) and we are grateful for computing time granted by the JARA Vergabegremium and provided on the JARA Partition part of the supercomputer CLAIX at RWTH Aachen University (project number jara0191).
This work was furthermore funded by the Deutsche Forschungsgemeinschaft (DFG, German Research Foundation) under Germany's Excellence Strategy – Cluster of Excellence Matter and Light for Quantum Computing (ML4Q) EXC 2004/1 – 390534769, and we thank the Bavarian Ministry of Economic Affairs, Regional Development and Energy for financial support within High-Tech Agenda Project “Bausteine f\"ur das Quantencomputing auf Basis topologischer Materialien mit experimentellen und theoretischen Ans\"atzen“.


\section*{Competing Interests}

The authors declare no competing financial or non-financial interests.

\appendix
\counterwithin{figure}{section}

\section{Supplement}

\subsection{Wave function localization}
\label{app:wfloc}

\begin{figure*}
    \centering
    \includegraphics[width=0.7\linewidth, trim={0cm 0.0cm 0cm 0cm}, clip]{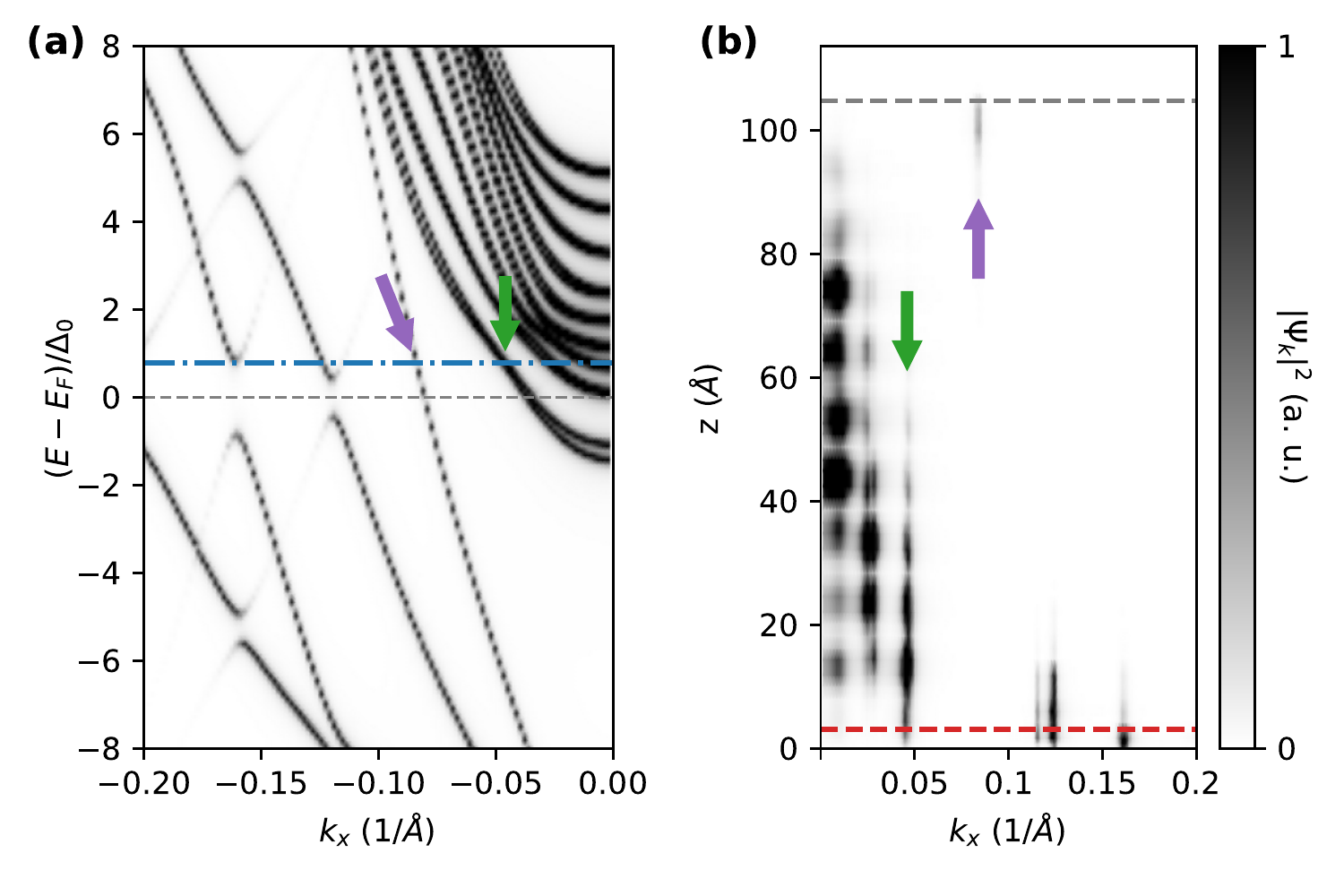}
    \caption{Superconducting band structure \textbf{(a)} and wave function localization \textbf{(b)} of 10QL Bi$_2$Te$_3$ proximitized by an interface with Nb(111). The band structure in (a) is integrated over the full heterostructure and the wave function localization in (b) is shown along the blue dash-dotted line in (a). The red and grey dashed lines in (b) separate the regions of the superconductor ($z<3\,\mathrm{\AA}$), the TI film ($3\,\mathrm{\AA} \le z \ge 105\,\mathrm{\AA}$) and the vacuum ($z>105\,\mathrm{\AA}$) in the heterostructure. The purple arrows at $k_x\approx0.08\,\mathrm{\AA}$ highlight the topological surface state on the free side. The green arrows at $k_x\approx0.04\,\mathrm{\AA}$ indicate the interface states.}
    \label{fig:wfloc}
\end{figure*}

Figure~\ref{fig:wfloc} shows the superconducting band structure of 10QL Bi$_2$Te$_3$ proximitized by an interface to Nb(111) which is compared to the localization of the wave functions throughout the heterostructure. We find that the states that originate from the Nb superconductor and penetrate only little into the TI region ($|k_x|>0.1\,\mathrm{\AA}$) have a large superconducting gap. The topological surface that is located on the free side (highlighted by purple arrows) on the other hand has no wave function overlap with the Nb region and therefore is not proximitized. The interface states that form due to charge-transfer induced band bending at the TI/Nb interface also only has a marginal extend of the wave function with the Nb region and therefore only picks up a tiny superconducting gap as discussed in the main text.

\bibliography{bibliography}

\end{document}